\documentclass[12pt]{article}
\usepackage{graphicx}
\usepackage{latexsym}
\usepackage{amsmath,amsfonts,amsthm}
\renewcommand{\arraystretch}{0.75}

\newcommand{\AGf}{\ensuremath{\mathcal{A}}}
\newcommand{\PGf}{\ensuremath{\mathcal{P}}}
\newcommand{\QGf}{\ensuremath{\mathcal{Q}}}
\newcommand{\TGf}{\ensuremath{\mathcal{T}}}
\newcommand{\RGf}{\ensuremath{\mathcal{R}}}

\begin{document}
\title{Enumerations of lattice animals and trees}
\author{Iwan Jensen\thanks{e-mail: I.Jensen@ms.unimelb.edu.au} \\
Department of Mathematics and Statistics, \\
The University of Melbourne,\\
Victoria 3010, Australia}
\date{\today}
\maketitle
\bibliographystyle{plain}
\begin{abstract}
We have developed an improved algorithm that allows us to enumerate the 
number of site animals on the square lattice up to size 46. 
We also calculate the number of lattice trees up to size 44  and 
the radius of gyration of both lattice animals and trees up to size 42.
Analysis of the resulting series yields an improved estimate,
$\lambda =  4.062570(8)$, for the growth constant of lattice animals,
and, $\lambda_0 =  3.795254(8)$, for the growth constant of trees,
and confirms to a very high degree of certainty that both the animal
and tree generating functions have a logarithmic divergence. 
Analysis of the radius of gyration series yields the estimate,
$\nu = 0.64115(5)$, for the size exponent.
\end{abstract}

{\bf KEY WORDS:} Lattice animals; Exact enumeration; Computer algorithms

\section{Introduction}

The enumeration of lattice animals is a classical combinatorial 
problem of great interest in it own right \cite{Golomb}.
Lattice animals are connected subgraphs of a lattice. A {\em site}
animal can be viewed as a finite set of lattice sites connected
by a network of nearest neighbor bonds. The fundamental problem
is the calculation (up to translation) of the number of animals, 
$a_n$, with $n$ sites. In the physics literature lattice animals are 
often called {\em clusters} due to their close relationship
to percolation problems \cite{Stauffer}. Series expansions 
for various percolation properties, such as the percolation probability 
or the average cluster size, can be obtained as weighted sums over 
the number of lattice animals, $g_{n,m}$, enumerated according to the 
number of sites $n$ and perimeter $m$ \cite{Domb,Sykes}. 
In  mathematics, and combinatorics in particular, the term 
polyominoes is frequently used. A polyomino is a set of lattice cells 
joined at their edges. So polyominoes are identical to site animals on the 
dual lattice. Furthermore, the enumeration of lattice animals has 
traditionally served as a benchmark for computer performance and algorithm 
design \cite{Lunnon}--\cite{Conway}. 

Lattice trees form a proper subset of lattice animals, and can be defined
as those animals containing no circuits. Another way of defining
trees is that a tree is a finite connected set of sites with the
property that a walk starting from any given site cannot return to
the original site without self-intersections. 
Lattice trees have been suggested as a model of branched 
polymers \cite{Lubensky}.
Lattice animals and trees are expected to belong to the same
universality class \cite{Lubensky,Duarte} and thus have the same
critical exponents.

An algorithm for the calculation of $g_{n,m}$ has been published by
Martin \cite{Martin} and Redner \cite{Redner}. It was used by Sykes
and co-workers to calculate series expansions for percolation problems
on various lattices. In particular Sykes and Glen \cite{Sykes} calculated
$g_{n,m}$ up to $n=19$ on the square lattice, and thus obtained the number 
of lattice animals, $a_n=\sum_m g_{n,m}$, to the same order. Redelmeier 
\cite{Redelmeier} presented an improved algorithm for the enumeration of 
lattice animals and extended the results to $n=24$. This algorithm was 
later used by Mertens \cite{Mertens90} to devise an improved algorithm for 
the calculation of $g_{n,m}$ and a parallel version of the algorithm 
appeared a few years later \cite{Mertens92}. The next major advance
was obtained by Conway \cite{Conway} who used the finite lattice
method with an associated transfer-matrix algorithm to calculate
$a_{n}$ and numerous other series  up to $n=25$ \cite{CG95}. 
In unpublished work Oliveira e Silva
\cite{Silva} used the parallel version of the Redelmeier algorithm
\cite{Mertens92} to extend the enumeration to $n=28$. In this work
we use an improved version of Conway's algorithm to extend the enumeration 
to $n=46$. We also calculate the number of lattice trees up to $n=44$ and 
the radius of gyration of lattice animals and trees up to $n=42$.

The quantities and functions we consider in this paper are: (i) the 
number of lattice animals $a_n$ and the associated generating function, 
$\AGf (u)= \sum a_n u^n$; (ii)  the number of lattice trees $t_n$ with
generating function, $\TGf (u)= \sum t_n u^n$; and (iii) the mean-square 
radius of gyration of animals or trees of size $n$, 
$\langle R^2 \rangle _n$. These quantities are expected to behave as

\begin{eqnarray}\label{eq:coefgrowth}
a_n & = & A \lambda^n n^{-\tau}[1+o(1)], \nonumber \\
t_n & = & T \lambda_0^n n^{-\tau}[1+o(1)], \\
\langle R^2 \rangle _n & = & R n^{2\nu}[1+o(1)], \nonumber
\end{eqnarray}
\noindent
where $\lambda$ and $\lambda_0$ are the reciprocals $u_c^{-1}$ of the critical 
point of, respectively, the animal and tree generating functions. 
From numerical evidence it is well-established that $\tau =1$. 

In Section~\ref{sec:enum} we give a detailed description of the finite 
lattice method for enumerating lattice animals. Some initial results of
the analysis of the series are presented in Section~\ref{sec:analysis}.

\section{Enumerations of lattice animals and trees \label{sec:enum}}

The method we use to enumerate site animals and trees on the square lattice 
is based on the method used by Conway \cite{Conway} for the calculation of 
series expansions for percolation problems, and is similar to methods
devised by Enting for enumeration of self-avoiding polygons \cite{Enting}
or the algorithm used by Derrida and De Seze to study percolation
and lattice animals \cite{Derrida}. 
In the following we give a detailed description of the algorithm used
to count lattice animals. We then show how to generalise the method to 
calculate the radius of gyration and obtain series for lattice trees.

\subsection{Transfer matrix algorithm}

The number of animals that span rectangles of width $W$ and length $L$ 
are counted using a transfer matrix 
algorithm. By combining the results for all $W\times L$ rectangles with 
$W \leq W_{\rm max}$ and $W+L \leq 2 W_{\rm max}+1$ we can count all animals 
up to $n=2 W_{\rm max}$. Due to symmetry we only consider rectangles 
with $L \geq W$ and thus count the contributions for rectangles with 
$L > W$ twice.

\begin{figure}
\begin{center}
\includegraphics[scale=0.8]{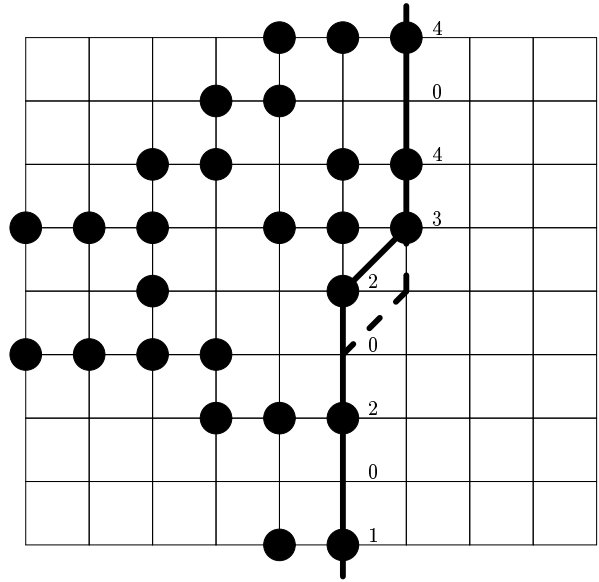}
\end{center}
\caption{\label{fig:transfer}
A snapshot of the intersection (solid line) during the transfer matrix 
calculation on the square lattice. Animals are enumerated by successive
moves of the kink in the boundary, as exemplified by the position given 
by the dashed line, so that one site at a time is added to the rectangle. 
To the left of the boundary we have drawn an example of a 
partially completed animal. Numbers along the boundary indicate the
encoding of this particular configuration.}
\end{figure}

The transfer matrix technique involves drawing a boundary line through the
rectangle intersecting a set of $W$ sites. For each configuration of 
occupied or empty sites along the boundary we maintain a generating function 
for partially completed animals intersecting the boundary in that particular 
pattern. Animals in a given rectangle are enumerated by moving the boundary 
so as to add one site at a time, as shown in figure~\ref{fig:transfer}. Each 
configuration can be represented by a set of states $\{\sigma_i\}$, where the 
value of the state $\sigma_i$ at position $i$ must indicate first of all 
if the site is occupied or empty. An empty site is simply indicated by 
$\sigma_i=0$. Since we have to ensure that we count only connected graphs
more information is required if a site is occupied. In short we need a way 
of describing which other occupied sites on the boundary it is connected 
to via a set of occupied sites to the left of the boundary. The most compact 
encoding of this connectivity is \cite{Conway}

\begin{equation}\label{eq:states}
\sigma_i  = \left\{ \begin{array}{rl}
0 &\;\;\; \mbox{empty site},  \\ 
1 &\;\;\; \mbox{occupied site not connected to others on the boundary}, \\
2 &\;\;\; \mbox{first among a set of connected boundary sites},\\ 
3 &\;\;\; \mbox{intermediate among a set of connected boundary sites}, \\
4 &\;\;\; \mbox{last among a set of connected boundary sites}.
\end{array} \right.
\end{equation}
\noindent
Configurations are read from the bottom to the top. As an example the 
configuration along the boundary of the partially completed animal in 
figure~\ref{fig:transfer} is $\{102023404\}$.

In addition to the configuration of states along the boundary line we 
also have to specify whether or not the partially completed animals 
include sites on the lower and/or upper borders of the rectangle. 
This can simply be done by marking a configuration with a 0 if
none of the borders have been touched, and a 1, 2 or 3 if, respectively,
the lower border, upper border or both borders have been touched. In this 
way we can be sure to count only those animals which span a given rectangle
in the vertical direction. That all animals span the horizontal 
direction is  ensured by the set updating rules detailed below.

The total configuration of occupied sites and the touching of the
borders can be encoded by a pair of integers $(S,k)$, where $k$ indicates 
which borders have been touched, and $S$ is the integer whose binary 
representation is obtained by assigning 3 bits to each $\sigma_i$ in
the configuration of occupied sites, $S=\sum_{i=0}^{W-1} \sigma_i 8^i$. 
We shall call such a $(S,k)$-pair a {\em signature}, and in practise
represent it by an integer $\widehat{S}=S+k*8^W$. For $W \leq 20$
a signature can thus conveniently be stored in the computer as a 
64-bit integer, while for $W>20$ we need to switch to a more
complicated representation, say, in terms of several 16-bit integers. 
Often we shall explicitly write out the 
configuration $\{\sigma_i\}$ instead of $S$ and use the notation
$\{S_1 S_2\}$ to indicate a configuration obtained by concatenating 
the strings $S_1$ and $S_2$. 

The major improvement of the method used to enumerate animals in this 
paper is that we require animals to span the rectangle in {\em both} 
directions. In the original approach \cite{Conway} animals were
only required to span in the lengthwise direction and animals of
width less than $W$ were generated many times.
It is however easy to obtain the animals of width exactly $W$ and
length exactly $L$ from this enumeration \cite{Enting}.
The only drawback of the new approach is that for most 
configurations we have to use four distinct generating functions. The 
major advantage is that the 
memory requirement of the algorithm is exponentially smaller. 

Realizing the full savings in memory usage comes from two enhancements to 
the original algorithm. Firstly, for each configuration we keep track of 
the current minimum number of occupied sites $N_{\rm cur}$ which have been 
inserted to the left of the intersection in order to build up that particular 
configuration. Secondly, we  calculate the minimum number of additional sites
$N_{\rm add}$ required to produce a valid animal. There are three 
contributions, namely the number of sites required to connect all the 
separate pieces of the partially completed animal, the number of sites 
needed to ensure that the animal touches both the lower and upper 
boundary, and finally the number of sites needed  to extend at 
least $W$ columns in the length-wise direction. If the sum 
$N_{\rm cur}+N_{\rm add} > 2W_{\rm max}$ we can discard the partial
generating function for that  configuration because it won't make a 
contribution to the animal count up to the size we are trying to obtain. 
Furthermore, for any $W$ we know that contributions will start at $2W-1$ 
since the smallest animals have to span a $W\times W$ rectangle. So for 
each configuration we need only retain $2(W_{\rm max}-W)+1$ terms of the 
generating functions. With the original algorithm contributions started 
at $W$ because the animals were required to span only in the length-wise 
direction. 

\subsubsection{Derivation of updating rules}

In Table~\ref{tab:update} we have listed the possible local `input' 
states and the `output' states which arise as the kink in the boundary
is propagated by one step.  The most important boundary
site is the `lower' one situated at the bottom of the kink (the site
marked with the second `2' in figure~\ref{fig:transfer}). This is
the position in which the lattice is being extended and obviously the
new site can be either empty or occupied. The second most
important boundary site is the `upper' one at the top of the kink
(the site marked `3' in figure~\ref{fig:transfer}). The state of
the upper site is very important in determining the state of the
lower site when occupied. The state of the upper site is likely 
to be changed as a result of the move. In addition the state of
a site further afield may have to be changed if a branch of
a partially completed animal terminates at the new site or if
two independent components of a partially completed animal join
at the new site. In the following we give the details
of how some of these updating rules are derived. We shall refer to
the signature before the the move as the `source' and a
signature produced as a result of the move as a `target'.

\begin{table}
\caption{\label{tab:update}
The various `input' states and the `output' states which arise as the 
boundary line is moved in order to include one more site of the lattice.
Each panel contains two `output' states where the left (right) most is the 
configuration in which the new site is empty (occupied).}
\begin{center}
\renewcommand{\arraystretch}{0.9}
\begin{tabular}{|c|cc|cc|cc|cc|cc|}  \hline  \hline
\raisebox{-1.5mm}{Lower}\raisebox{-0.5mm}{\Large $\backslash$}\raisebox{0.5mm}{Upper} 
 &\multicolumn{2}{c|}{0} &\multicolumn{2}{c|}{1} & \multicolumn{2}{c|}{2}
& \multicolumn{2}{c|}{3} & \multicolumn{2}{c|}{4} \\ 
\hline
0  & $00$   & $10$   & $01$   & $24$   & $02$   & $23$
   & $03$   & $33$   & $04$   & $34$\\    \hline
1  & add   & $10$   & $--$   & $24$   & $--$   & $23$
   & $--$   & $33$  & & \\    \hline
2  & $\overline{00}$   &$20$   & $\overline{01}$   & $23$
   & $\overline{02}$   & $\widehat{23}$
   & $02$   & $23$   & $01$   & $24$ \\   \hline
3  & $00$   & $30$   & $01$   & $33$   & $02$   & $ \widehat{33}$
   & $03$   & $33$   & $04$   & $34$\\    \hline
4  & $\overline{00}$   & $40$   & $\overline{01}$   & $34$
   & $\overline{02}$ & $33$ 
   & $\overline{03}$ & $\widehat{33}$ & &\\ 
   \hline \hline
\end{tabular}
\end{center}
\end{table}

\begin{description}

\item{00:} The lower and upper sites are empty. If 
the new site is empty the signature is unchanged. If
the new site is occupied it isn't connected to other
sites in the boundary and is in state 1. 
From the source configuration $\{S_1 00 S_2\}$ we get
the targets  $\{S_1 00 S_2\}$ and $\{S_1 10 S_2\}$.

\item{01:} The lower site is empty and the upper site is isolated. 
If the new site is empty the signature is unchanged. If
the new site is occupied it is connected to the upper site 
and is in state 2 while the state of the upper
site is changed to state 4.

\item{02:} When the new site is occupied it is connected to the upper 
site. The state of the lower site becomes 2 (the new first site in
the set) while the state of the upper site is changed to 3
(it is now an intermediate site).

\item{10:} The lower site was an isolated occupied site
so if the new site is empty we have created a separate graph. 
This is only allowed if there are no other
occupied sites on the boundary line (otherwise we generate
graphs with separate components) and if both the lower and
upper borders have been touched. The result are
valid lattice animals. The generating function is accumulated 
into the final animal generating function. 
If the new site is occupied it isn't connected to other
sites in the boundary and is therefore still in state 1.

\item{11:} The new site has to be occupied and it is connected to the 
upper site. The new site is in state 2 while the state of the upper
site is changed to state 4. 

\item{14:} This situation never occur. The upper site is
the last among a set of occupied sites.
This implies that the site immediately to the left of the upper
site is occupied, this in turn is connected to the lower
site, which therefore cannot be an isolated occupied site. 

\item{20:} The lower site is the first among a set of occupied 
sites, so if the new site is empty, another site in this set
changes its state. Either the {\em first} intermediate site
becomes the new first site, and its state is changed from 
3 to 2, or, if there are no intermediate sites, the last site
becomes an isolated occupied site, and its state is changed
from 4 to 1. Note that there could be connected parts
of the animal interspersed between the first site and
the matching intermediate or last site, so locating the
site which has to be changed requires a little computation.
This is illustrated in figure~\ref{fig:transfer} where
the first 2 is connected to the last 4, and a piece of the
animal is placed in between these two sites. In this example
if the first 2 became a 0 the last 4 becomes a 1, while if
the second 2 becomes a 0 the 3 above it becomes a 2. In general
the nesting can be quite complicated and the general rule for
updating the configuration is as follows: Start from the 2,
which we are changing to a 0, and move upwards in the configuration.
Count the number of 2's and 4's as we pass them. If an equal
number has been passed and we encounter a 3 or 4 this is the
matching site we are looking for and it is changed either
to 2 or 1. This change of a matching site is indicated in 
Table~\ref{tab:update} by over-lining. When the
new site is occupied the configuration is unchanged. So
from the source  $\{S_1 20 S_2\}$ we
get the targets  $\{S_1 00 \overline{S_2}\}$ and  $\{S_1 20 S_2\}$.

\item{22:} The updating when the new site is empty is as before.
When the new site is occupied the connectivity is altered since
we are joining two separate pieces of the animal. The new site
remains the first site in the joined piece while the upper
site becomes an intermediate site. The last site
in the set of connected sites starting at the upper site also
becomes an intermediate site in the joined piece. Locating this
site is similar to the operation indicated by over-lining.
However, in this case we ignore sites in state 3 and the matching
site in state 4 becomes a 3. We indicate this type of transformation
by putting a hat over the string. The source  
$\{S_1 22 S_2\}$ gives rise to the targets $\{S_1 02 \overline{S_2}\}$ 
and  $\{S_1 23 \widehat{S_2}\}$.

\item{40:} When the new site is empty we must change a matching
site, either an intermediate site to a last site or a first site to
an isolated occupied site. The transformation is similar to the case
20, but we have to search downwards in the configuration. 

\item{43:} When the new site is occupied we change the connectivity.
The first site, from the set of sites
connected to the lower site, is changed to an intermediate site.
This transformation is similar to the `hat' transformation described 
at case 22, but we now have to search downwards
in the configuration. 

\item{44:} This can't happen for the same reason that 14 is impossible. 

\end{description}

\subsubsection{The algorithm}

As a new site is added to the lattice we construct a new set of
partial generating functions from the existing set. This can be
done by running through all members of the existing set.
Using bit-masking we can extract the states of the lower and
upper sites and then apply the relevant updating rules, which
generate at most 2 target signatures. First
we check if the signature already exists, if so the generating
functions of the source and target are added (with an addition
weight factor $u$ on the source if the new site is occupied).
If the signature doesn't exist already, we check whether
or not it makes a contribution, that is, we see if 
$N_{\rm cur}+N_{\rm add} \leq 2W_{\rm max}$ ($N_{\rm cur}$ of
the target is $N_{\rm cur}$ of the source if the new site is
empty and $N_{\rm cur}+1$ otherwise). If the target makes a
contribution it is assigned a storage position and its 
generating function is the generating function of the source
(again with an extra factor of $u$ if the new site is occupied). 
When the target generating functions have been created the
storage position of the source generating function is released  
since it is no longer required and thus can be recycled.

The algorithm for the enumeration of animals spanning a $W\times L$ 
strip is:

\begin{enumerate}

\item Start by inserting an isolated occupied site in the 
top left corner. This configuration has the signature $(8^{W-1},2)$, 
which enters with a count of 1.

\item For $j$ from 2 to $W-1$ add a site to the lattice in the
first column. Run through all existing signatures using the updating 
rules described above (note that as this is the first column the
lower site is always empty). Add an additional configuration
with a single occupied site at position $W-j$ with a count of 1. 
These configurations have the signature  $(8^{W-j},0)$, since none 
of the borders have been touched.

\item Put in the last site in the first column. Again we run
through all existing signatures. If the new site occupied we
have to mark the signature as having touched the lower border.
Add an additional configuration with a single occupied site in
the lower left corner with a count of 1, the signature is $(1,1)$.

\item Put in the top site in the next column. Run through all existing 
signatures. Since we are at the top border we only use the updating 
rules in Table~\ref{tab:update} with the upper site in state 0
(obviously the lower site cannot be in states 2 or 3). If the new
site is occupied make sure that the signature is marked as having
touched the upper border. In this generic case we do {\em not}
put in the additional configuration of a single isolated occupied
site since it would not touch the left-most border.

\item For $j$ from 2 to $W-1$ add a site to the lattice in row $W-j$.
Run through all existing signatures using the updating rules. Again
no isolated occupied should be inserted.

\item Put in the last site in the column. If the new
site is occupied make sure that the signature is marked as having
touched the lower border. 

\item If the number of completed columns is less than $L$ go to 4.

\end{enumerate}

\subsubsection{Computational complexity}

\begin{figure}
\begin{center}
\includegraphics[scale=0.6]{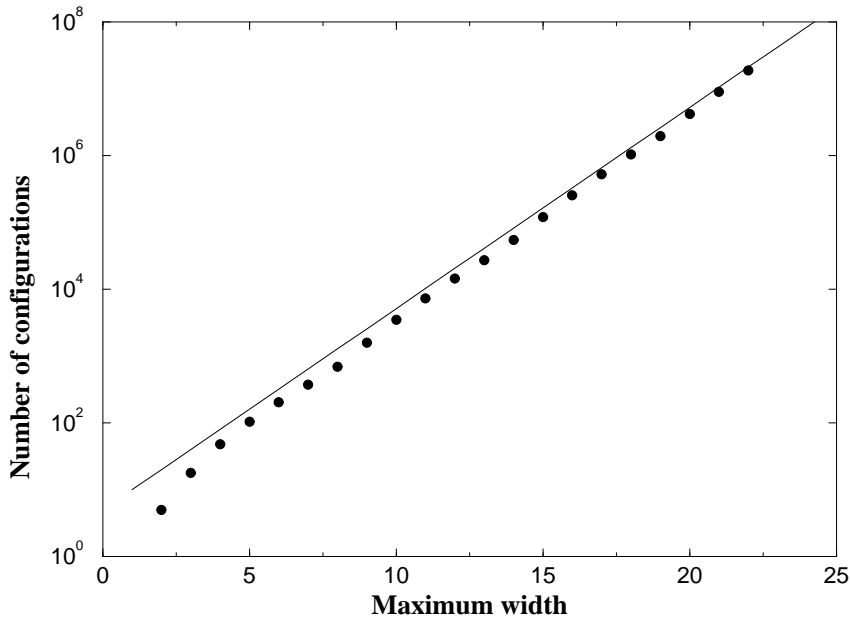}
\end{center}
\caption{\label{fig:memuse}
The number of configurations required in order to count the
number of lattice animals correct up to twice the maximum width.
The solid line is drawn as a guide to the eye and would correspond
to a growth rate of exactly 2.} 
\end{figure}

The algorithm has exponential complexity, that is the time required
to obtain the animals up to size $n$ grows exponentially with $n$. Time 
and memory requirements are basically proportional to the maximum number 
of distinct configuration generated during a calculation. This in turn 
depends on the maximum number of terms we wish to calculate and thus on 
$W_{\rm max}$. In figure~\ref{fig:memuse} we have shown how the maximal 
number of configurations, $N_{\rm Conf}$, grows with $W_{\rm max}$. From 
this it is clear $N_{\rm Conf} \propto a^{W_{\rm max}}$, and from
the figure we estimate that $a$ is a little larger than 2. Since we 
obtain $2W_{\rm max}$ terms the computational complexity grows 
exponentially with growth constant $\sqrt{a}$. Note that this is much
better than a direct enumeration in which time requirements are 
proportional to the number of animals and therefore has the growth
constant, $\lambda \simeq 4.06\ldots$, of lattice animals. The price
we have to pay for a faster algorithm is that the memory requirement also
grows exponentially like $N_{\rm Conf}$, whereas in direct enumerations
the memory requirement typically grows like a polynomial in the number 
of terms.

\subsubsection{Further particulars}

Finally a few remarks of a more technical nature. The number of contributing 
configurations becomes very sparse in the total set of possible states along 
the boundary line and as is standard in such cases one uses a hash-addressing 
scheme \cite{Mehlhorn}. Since the integer coefficients occurring in the  
expansion become very large, the calculation was performed using modular 
arithmetic \cite{Knuth}. This involves performing the calculation modulo 
various prime numbers $p_i$ and then reconstructing the full integer
coefficients at the end. In order to save memory we used primes of the form 
$p_i=2^{15}-r_i$ so that the residues of the coefficients in the polynomials 
could be stored using 16 bit integers. The Chinese remainder theorem
ensures that any integer has a unique representation in terms of residues. If 
the largest integer occurring in the final expansion is $m$, then we 
have to use a number of primes $k$ such that $p_1p_2\cdots p_k > m$.  Up to 
6 primes were needed to represent the coefficients correctly.

\subsection{Calculation of the radius of gyration}

In the following we show how the definition of the radius of gyration 
can be expressed in a form suitable for a transfer matrix calculation.
As is well-known the radius of gyration of $n$ points at positions
${\bf r}_i$ is 

\begin{equation} \label{eq:rg}
n^2 R^2_n = \sum_{i>j} ({\bf r}_i-{\bf r}_j)^2 =
(n\!-\!1)\sum_i (x_i^2+y_i^2)-2\sum_{i>j}(x_ix_j+y_iy_j).
\end{equation}

This last expression is suitable for a transfer matrix calculation. 
As usual we actually calculate the generating function,
$\RGf^2_g(u) = \sum_n a_n\langle R^2 \rangle _n n^2 u^n$, since this
ensures that the coefficients are integers. In order to
do this we have to maintain five partial generating functions
for each signature, namely

\begin{itemize}
\item $A(u)$, the number of (partially completed) animals.
\item $R^2(u)$, the sum over animals of the squared components of the 
distance vectors.
\item $X(u)$, the sum of the $x$-component of the distance vectors.
\item $Y(u)$, the sum of the $y$-component of the distance vectors.
\item $XY(u)$, the sum of the `cross' product of the components of the  
distance vectors, e.g., $\sum_{i>j}(x_ix_j+y_iy_j)$.
\end{itemize}

As the boundary line is moved to a new position each configuration $S$ 
might be generated from several configurations $S'$ 
in the previous boundary position. The partial generation functions are
updated as follows

\begin{eqnarray}
A(u,S) & = & \sum_{S'} 
  u^{n(S')} A(u,S'), \nonumber \\
R^2(u,S) & = & \sum_{S'}  
  u^{n(S')}[R^2(u,S')+n(S')(x^2+y^2)A(u,S')],\nonumber \\ 
X(u,S) & = & \sum_{S'}   
  u^{n(S')}[X(u,S')+  x n(S')A(u,S')], \\ 
Y(u,S) & = & \sum_{S'}   
  u^{n(S')}[Y(u,S')+ y n(S') A(u,S')], \nonumber \\ 
XY(u,S) & = & \sum_{S'}  
  u^{n(S')} [XY(u,S')+ xn(S')X(u,S')+yn(S')Y(u,S')] \nonumber 
\end{eqnarray}
\noindent
where $n(S')$ is the number of occupied site added to the 
animal.

\subsection{Enumeration of lattice trees}

Lattice trees can be enumerated in essentially the same manner
as animals. We merely get some further restrictions on the 
rules listed in Table~\ref{tab:update}. The necessary restriction is
that the new site cannot be occupied if the lower and upper sites
already are connected, since this would obviously result in the formation
of a circuit. So in the cases `23', `24', `33', and `34' the new
site cannot be occupied, otherwise the updating rules are identical to
those for animals. 

\section{Analysis of the series \label{sec:analysis}}

The series listed in Table~\ref{tab:series} have coefficients which 
grow exponentially, with sub-dominant term given by a critical exponent.
The generic behaviour is $g_n \sim \mu^n n^{\xi-1}$, and hence the
generating function has the behaviour,
$G(u) =\sum_n g_n u^n \sim (1-u/u_c)^{-\xi},$ 
where $u_c=1/\mu$.  From (\ref{eq:coefgrowth}) we get 
the following predictions for the animal generating functions:

\begin{eqnarray}\label{eq:genfunc}
\AGf (u)& = &\sum_n a_n u^n = A(u)(1-u\lambda)^{1-\tau}, \\
\RGf^2_g (u)& = &\sum_n a_{n}\langle R^2 \rangle _{n}n^2 u^n =
    \sum_{n} r_n u^n \sim R(u)(1-u\lambda)^{-(\tau+2\nu+1)}.
\end{eqnarray}
\noindent
Similar expressions hold for the corresponding
tree generating functions though with a different growth constant
$\lambda_0$. So the animal and tree generating functions are
expected to have a logarithmic singularity, while the radius of
gyration series are expected to diverge with an exponent $2+2\nu$,
where we assumed the conjecture $\tau=1$ to be correct. 

\begin{table}
\caption{\label{tab:series}
The number of lattice animals, $a_n$, lattice trees, $t_n$, and the 
coefficients in the respective generating functions for their radius 
of radius.}
\begin{center}
\tiny
\renewcommand{\arraystretch}{0.85}
\begin{tabular}{rllll}  \hline  \hline
$n$  & $a_n$ & $n^2 a_n \langle R^2 \rangle _n$ &
       $t_n$ & $n^2 t_n \langle R^2 \rangle _n$ \\
\hline
1 & 1 & 
  & 
1 & 
  \\ 
2 & 2 & 
2 & 
2 & 
2 \\ 
3 & 6 & 
28 & 
6 & 
28 \\ 
4 & 19 & 
252 & 
18 & 
244 \\ 
5 & 63 & 
1840 & 
55 & 
1680 \\ 
6 & 216 & 
11924 & 
174 & 
10214 \\ 
7 & 760 & 
71476 & 
570 & 
57476 \\ 
8 & 2725 & 
405204 & 
1908 & 
305476 \\ 
9 & 9910 & 
2202724 & 
6473 & 
1553632 \\ 
10 & 36446 & 
11590162 & 
22202 & 
7641218 \\ 
11 & 135268 & 
59417180 & 
76886 & 
36608932 \\ 
12 & 505861 & 
298186524 & 
268352 & 
171666468 \\ 
13 & 1903890 & 
1470151308 & 
942651 & 
790650724 \\ 
14 & 7204874 & 
7140410208 & 
3329608 & 
3586822020 \\ 
15 & 27394666 & 
34237750548 & 
11817582 & 
16062938368 \\ 
16 & 104592937 & 
162350915772 & 
42120340 & 
71135451440 \\ 
17 & 400795844 & 
762391407024 & 
150682450 & 
311964025352 \\ 
18 & 1540820542 & 
3549556185044 & 
540832274 & 
1356392904818 \\ 
19 & 5940738676 & 
16400558514664 & 
1946892842 & 
5852609697844 \\ 
20 & 22964779660 & 
75263022053196 & 
7027047848 & 
25081266854732 \\ 
21 & 88983512783 & 
343273594201564 & 
25424079339 & 
106827845665800 \\ 
22 & 345532572678 & 
1557003525653380 & 
92185846608 & 
452491861285360 \\ 
23 & 1344372335524 & 
7026663432447428 & 
334925007128 & 
1906994132045328 \\ 
24 & 5239988770268 & 
31565321263960648 & 
1219054432490 & 
8000039128666316 \\ 
25 & 20457802016011 & 
141201716724567204 & 
4444545298879 & 
33420021839691568 \\ 
26 & 79992676367108 & 
629195375725422292 & 
16229462702152 & 
139072296450104904 \\ 
27 & 313224032098244 & 
2793681657766773944 & 
59347661054364 & 
576665646646645628 \\ 
28 & 1228088671826973 & 
12363167055143142440 & 
217310732774774 & 
2383267493411599452 \\ 
29 & 4820975409710116 & 
54544020640717162468 & 
796703824808133 & 
9819513412114987172 \\ 
30 & 18946775782611174 & 
239950473304391505440 & 
2924252282840112 & 
40342989684066501360 \\ 
31 & 74541651404935148 & 
1052776828941036051656 & 
10744903452821876 & 
165306633582934256304 \\ 
32 & 293560133910477776 & 
4607511085613062500648 & 
39521236485358584 & 
675665329410485731504 \\ 
33 & 1157186142148293638 & 
20117772038497717315976 & 
145503056229823138 & 
2755244324874079014600 \\ 
34 & 4565553929115769162 & 
87647322688578954475976 & 
536170499427125956 & 
11210822859036572606668 \\ 
35 & 18027932215016128134 & 
381065022045089903130608 & 
1977427804277385532 & 
45521864027574363668480 \\ 
36 & 71242712815411950635 & 
1653532426475382203248376 & 
7298688919041663694 & 
184484204103594541676168 \\ 
37 & 281746550485032531911 & 
7161875592952535220704656 & 
26959808299736689704 & 
746280442016872847892140 \\ 
38 & 1115021869572604692100 & 
30965967036768698515049964 & 
99655022360008737496 & 
3013643917345287146830452 \\ 
39 & 4415695134978868448596 & 
133667644427251173600540220 & 
368617606804069356072 & 
12149786877969635612633264 \\ 
40 & 17498111172838312982542 & 
576087681668533750775182764 & 
1364371688078200595674 & 
48906771633330499596166064 \\ 
41 & 69381900728932743048483 & 
2479166130662936965224977368 & 
5053070869464350119408 & 
196574389975234157470737780 \\ 
42 & 275265412856343074274146 & 
10653909826486480285867012570 & 
18725415026570087447460 & 
788994500989152614915884776 \\ 
43 & 1092687308874612006972082 & 
  & 
69430306096976372288324 & 
  \\ 
44 & 4339784013643393384603906 & 
  & 
257571182441471056810356 & 
  \\ 
45 & 17244800728846724289191074 & 
  & 
  & 
  \\ 
46 & 68557762666345165410168738 & 
  & 
  & 
  \\ 

\hline \hline
\end{tabular}
\end{center}
\end{table}

\begin{table}
\caption{\label{tab:analysis} Estimates for the critical point $u_c$ and 
exponents $1-\tau$ and $1+\tau+2\nu$ obtained from second order 
inhomogeneous differential approximants to the series for the generating 
functions of lattice animals, lattice trees and their radius of gyration. 
$L$ is the order of the inhomogeneous polynomial.}
\begin{center}
\begin{tabular}{rllll}
\hline \hline
\multicolumn{5}{c}{Square lattice site animals} \\ \hline
\multicolumn{1}{c}{L} &
\multicolumn{1}{c}{$u_c$} &
\multicolumn{1}{c}{$1-\tau$} &
\multicolumn{1}{c}{$u_c$} &
\multicolumn{1}{c}{$1+\tau+2\nu$} \\ 
\hline 
 0  & 0.246149987(43)& -0.000523(46)                
    & 0.246150539(87)& 3.28413(11) \\         
 2 & 0.24614992(14)& -0.00043(14)       
    & 0.24615046(10)& 3.28402(28) \\         
 4 & 0.24615007(15)& -0.00055(16)        
   & 0.24615037(22)& 3.28394(30) \\         
 6 & 0.24614999(24)& -0.00046(25)        
   & 0.24615068(16)& 3.28426(22) \\         
 8 & 0.24615001(15)& -0.00052(13)         
   & 0.24615067(25)& 3.28432(44) \\         
10 & 0.24614997(22)& -0.00044(28)    
   & 0.24615055(31)& 3.28417(56) \\   
\hline \hline      
\multicolumn{5}{c}{Square lattice site trees} \\ \hline
\multicolumn{1}{c}{L} &
\multicolumn{1}{c}{$u_c$} &
\multicolumn{1}{c}{$1-\tau$} &
\multicolumn{1}{c}{$u_c$} &
\multicolumn{1}{c}{$1+\tau+2\nu$} \\ 
\hline 
 0 & 0.26348751(73)& -0.00039(55)         
   & 0.263487100(52)& 3.282325(42) \\         
 2 & 0.26348716(21)& -0.00014(13)         
   & 0.263487033(57)& 3.282276(35) \\         
 4 & 0.26348698(32)& 0.00002(27)         
   & 0.263487029(58)& 3.282276(35) \\         
 6 & 0.26348693(20)& 0.00000(12)        
   & 0.263487079(17)& 3.282308(12) \\         
 8 & 0.263486943(70)& 0.000009(58)          
   & 0.263487061(32)& 3.282297(20) \\         
 10 & 0.26348700(17)& -0.00001(10)         
    & 0.263487059(19)& 3.282296(13) \\
\hline \hline        
\end{tabular}
\end{center}
\end{table}

In the first stage of the analysis, we used the method of differential 
approximants \cite{Guttmann89}. Estimates of the critical point and critical 
exponent were obtained by averaging values obtained from second order 
inhomogeneous differential approximants.   
In Table~\ref{tab:analysis} we have listed the estimates obtained from 
this analysis. The error quoted for these estimates reflects the spread 
(basically one standard deviation) among the approximants. Note that these 
error bounds should {\em not} be viewed as a measure of the true error as 
they cannot include possible systematic sources of error. From this
we see that the animal generating function has a singularity at 
$u_c=0.246150(1)$, and thus we obtain the estimate, $\lambda =  4.06256(2)$,
for the growth constant. The exponent estimates are consistent with the 
expected logarithmic divergence, thus confirming the conjecture $\tau=1$.
The central estimates of $u_c$ obtained from the radius of gyration series
are a little larger than, but nonetheless consistent with the animal
generating function. From this analysis we see that the series has a 
divergence at $u_c$ with an exponent $2+2\nu = 3.2840(8)$, and thus
$\nu=0.6420(4)$.

The tree generating function has a singularity at $u_c=0.2634870(5)$,
and thus $\lambda_0 = 3.795254(8)$, with the expected logarithmic divergence. 
In this case estimates from the radius of gyration series yield 
$2+2\nu = 3.2823(1)$, and thus $\nu=0.64115(5)$. Since the $u_c$ estimates
from the two tree series are in excellent agreement we claim that
the best estimate for $\nu$ is the one obtained from the tree series.
This estimate is 
consistent with, but much more accurate than, the recent estimate 
$\nu=0.642(2)$ obtained from Monte Carlo simulations of lattice
trees \cite{You+Rensburg}. It is also consistent with the estimate
$\nu =0.6408(3)$ obtained using phenomenological renormalization
to lattice animals \cite{Derrida}.

A more detailed analysis of the animal series
was performed in \cite{JG2000}. It showed that in a plot of  
exponent vs  $u_c$ estimates, as  $1-\tau$ 
approach 0, $u_c$ approach 0.2461497. From the spread among the 
approximants we obtained the final estimate $u_c=0.2461496(5)$, and thus
the growth constant $\lambda = 4.062570(8)$. A similar analysis
of the radius of gyration series yielded estimates of $\nu$ consistent
with those obtained for lattice trees.

Finally we use the series to derive improved rigorous lower
bounds for the growth constants of lattice animals and trees.
Using concatenation arguments Rands and Welsh \cite{Rands} showed
that if we define a sequence $p_n$ such that

\begin{equation}
a_{n+1}=p_{n+1}+ p_{n} a_{2}+\ldots p_3 a_{n-1}+p_2 a_n,
\end{equation}
\noindent
and construct the generating functions 
\begin{equation}
\AGf^* (u) = 1 + \sum_{n=1}^{\infty} a_{n+1}u^n
\end{equation}
\noindent
and
\begin{equation}
\PGf (u) = \sum_{n=1}^{\infty} p_{n+1}u^n
\end{equation}
\noindent
then
\begin{equation}
\AGf^* (u) = 1+ \AGf^* (u)\PGf (u)
\end{equation}
\noindent
and  $\AGf^* (u)$ is singular when $\PGf (u)=1$. The coefficients in
$\PGf (u)$ are obviously known correctly to the same order $N=2W_{\rm max}-1$
as $\AGf^* (u)$. If we look at the polynomial $P_N$ obtained by truncating
$\PGf (u)$ at order $N$ then the unique positive zero, $1/\lambda_N$, of
$P_N-1=0$ is a lower bound for $\lambda$. Using our extended series we
find that $\lambda \geq 3.903184\ldots$.

For site trees the best lower bound appears to arise from a
different concatenation procedure \cite{Whittington}, which leads
to the equation

\begin{equation}
\TGf (u) = \frac{1- \QGf (u)}{1-2\QGf (u)}
\end{equation}
\noindent
and $\TGf (u)$ is singular when $\QGf (u) =1/2$. This approach yields
a lower bound for site trees, $\lambda_0 \geq 3.613957\ldots$.

\section{Conclusion}

We have presented an improved algorithm for the enumeration of site animals
on the square lattice. The computational complexity of the algorithm
is exponential with time (and memory) growing as $a^{n/2}$, where $a$  
appears to be a little larger than 2. Implementing this algorithm has 
allowed us to count the number of site animals up size 46. Our extended 
series enables us to give an improved estimate for the growth constant 
and confirm to a very high degree of certainty that the associated 
generating function has a logarithmic divergence. The algorithm was also 
modified to enumerate lattice trees up to size 44, and a generalised 
version was used to calculate the radius of gyration of animals and 
trees up to size 42. Analysis of the series confirmed that animals and 
trees belong to the same universality class and an accurate estimate
was obtained for the size exponent $\nu$.

\section*{E-mail or WWW retrieval of series}

The series for the generating functions studied in this paper 
can be obtained via e-mail by sending a request to 
I.Jensen@ms.unimelb.edu.au or via the world wide web on the URL
http://www.ms.unimelb.edu.au/\~{ }iwan/ by following the instructions.

\section*{Acknowledgements}

Financial support from the Australian Research Council is 
gratefully acknowledged.

\end{document}